# Intersection Two-Vehicle Crash Scenario Specification for Automated Vehicle Safety Evaluation Using Sequence Analysis and Bayesian Networks

[Preprint]*

Yu Song[1], Madhav V. Chitturi[2], David A. Noyce[2]

1. University of Connecticut, 2. University of Wisconsin-Madison

**Abstract**

This paper develops a test scenario specification procedure using crash sequence analysis and Bayesian network modeling. Intersection two-vehicle crash data was obtained from the 2016-2018 National Highway Traffic Safety Administration (NHTSA) Crash Report Sampling System (CRSS) database. Vehicles involved in the crashes are specifically renumbered based on their initial positions and trajectories. Crash sequences are encoded to include detailed pre-crash events and concise collision events. Based on sequence patterns, the crashes are characterized as 55 types. A Bayesian network model is developed to depict the interrelationships among crash sequence types, crash outcomes, human factors, and environmental conditions. Scenarios are specified by querying the Bayesian network's conditional probability table. Distributions of operational design domain (ODD) attributes (e.g., driver behavior, weather, lighting condition, intersection geometry, traffic control device) are specified based on conditions of sequence types. Also, distribution of sequence types is specified on specific crash outcomes or combinations of ODD attributes.

# 1 Introduction

Scenario-based testing is an essential part of automated vehicle (AV) safety evaluation, and generating challenging scenarios is critical for such testing (Koopman & Wagner, 2016; Riedmaier et al., 2020). Historical crash data consists of challenging scenarios faced by human drivers and is a good source of data for developing test scenarios that may also be challenging for AVs (Najm et al., 2007; Nitsche et al., 2017; Scanlon et al., 2021). As advanced driving assistance systems (ADAS) and automated driving systems (ADS) are developed to replace human drivers partially or fully, it is reasonable to expect AVs to have the capabilities to handle challenging scenarios faced by human drivers and mitigate crash outcomes. Therefore, by mining data of crashes involving human-driven vehicles, we would be able to generate safety-critical scenarios to test the capabilities of AVs in handling interactions with human-driven vehicles. The same approach can also be used to extract safety-critical scenarios from data of AV-involved crashes and generate scenarios to test AV interactions with other AVs (Song, 2021; Song et al., 2021).

Prior efforts in developing test scenarios using historical crash data have developed characterization of crashes to be used as representative scenarios for the evaluation of ADAS or ADS. The crash characterization was developed by summarizing and mining patterns in crash attributes (Najm et al., 2007; Nitsche et al., 2017; Sander & Lubbe, 2018; Sui et al., 2019; Watanabe et al., 2019; Esenturk et al., 2021, 2022). The end products from prior efforts – representative scenarios – lack considerations of crash progression, dynamics, and mechanisms, which are important information to distinguish crashes and their outcomes (Song et al., 2021; Wu et al., 2016).

The objective of this paper is to propose a crash-data-based scenario generating procedure that improves crash characterization and scenario specification by employing crash sequence analysis and Bayesian network modeling. Prior studies have defined scenario as a sequence of scenes, which include actions and interactions between moving objects, as well as the surrounding environment (Ulbrich et al., 2015). Layered models of potentially needed elements for scenario specification has been proposed by AV safety evaluation projects such as PEGASUS, which has a layer of moving object actions and interactions, and several other layers for elements in the surrounding environment (Sauerbier et al., 2019). In this study, a crash scenario is defined as *crash sequence + description of the operational design domain (ODD)*. Crash sequence analysis generates informative crash sequence types that describe crash dynamics and progression. Bayesian network modeling provides a model of crash mechanisms. The two analyses together enable the specification of crash scenarios depicted by the actions and interactions of moving participants, crash outcomes, and ODD variables including physical environmental conditions and human factors.

In this paper, we focus on applying the scenario generating procedure to intersection two-vehicle crashes. The reasons are: 1) intersection two-vehicle crashes are prevalent, making up more than a quarter of all crashes on United States roadways, and 2) intersection two-vehicle crashes contain complex interactions (both vehicle-vehicle and vehicle-environment) that need to be tested for AV safety proving. Crash data was obtained from the 2016-2018 Crash Report Sampling System (CRSS) of the United States National Highway Traffic Safety Administration (NHTSA). Following this section: Section 2: A literature review summarizing key literature in scenario generation using historical crash data, crash sequence analysis, and application of Bayesian networks in crash analysis. Section 3: Description of the CRSS data, special data preparation for scenario development, crash sequence data processing techniques, and other crash attributes used in modeling. Section 4: Explanation of methods employed in the scenario generating procedure. Section 5: Results of crash characterization based on sequences, Bayesian network modeling of variable dependencies, and a demonstration of scenario specification process. Section 6: Discussion and conclusion.

# 2 Literature Review
## 2.1 Scenario Generation Using Historical Crash Data

Historical crash data has been used to generate scenarios for different purposes, such as general vehicle safety testing, evaluation of crash avoidance systems, ADAS, and ADS (Najm et al., 2007; Aust, 2010; Kusano

& Gabler, 2012, 2013; Lenard et al., 2014; MacAlister & Zuby, 2015; Nitsche et al., 2017; Sander & Lubbe, 2018; Sui et al., 2019; Watanabe et al., 2019; Scanlon et al., 2021). Some prior research developed a comprehensive set of scenarios that cover as many crash types and ODDs as possible, using a national-level crash database and by summarizing crash attributes (Najm et al., 2007). Most prior studies focused on some specific types of crashes and extracted or characterized crash scenarios using national-level crash databases (Kusano & Gabler, 2012, 2013; Lenard et al., 2014; MacAlister & Zuby, 2015; Nitsche et al., 2017; Sander & Lubbe, 2018; Sui et al., 2019). Several studies proposed systematic methodologies to characterize crashes and specify representative scenarios (Aust, 2010; Nitsche et al., 2017; Watanabe et al., 2019). A most recent Waymo LLC study used municipal-level crash report census data to reconstruct crash scenarios in simulations for AV safety evaluation (Scanlon et al., 2021).

## 2.2 Crash Sequence Analysis

Crash sequences are chronologically ordered events that happened during the pre-crash and crash periods. Crash sequence analysis has a similar purpose with sequence analysis in biological and sociological research, focusing on identifying representative components of sequences, the similarity (or dissimilarity) between sequences, and the relationship between sequences and potential outcomes (Wu et al., 2016).

Prior studies on crash sequence analysis are limited. Krull et al. found significant relationships between the order of rollover and fixed-object collision events and single-vehicle crash injury severity outcomes (Krull et al., 2000). Wu et al. adapted sequence analysis methods from biological and sociological research to fatal single-vehicle run-off-road crash data, characterized crashes and found significant relationship between crash sequence types and crash injury severity outcomes (Wu et al., 2016). Song et al. used sequence analysis to characterize California AV crash sequences, found significant association between crash sequence types and multiple crash attributes, and proposed a scenario-based AV evaluation framework with crash sequences as the core component (Song et al., 2021). In more recent research, a methodology was developed for crash sequence analysis that could help select optimal techniques (e.g., sequence encodings, dissimilarity measures) for multiple different use cases in studying crashes (Song, 2021; Song et al., 2022). A case study of interstate single-vehicle crash characterization was used to demonstrate the effectiveness and usefulness of the proposed methodology.

## 2.3 Bayesian Networks for Crash Analysis

Bayesian networks are graphical models known for their use in revealing variable dependencies, and are widely applied in artificial intelligence, medical, and genetic research (Pearl, 2009). Prior applications of Bayesian networks in traffic crash research focused on crash forensics, estimating effects of contributing factors on crash outcomes, and predicting crash outcomes (Davis, 2001, 2003; Davis et al., 2011; de Oña et al., 2013; Ma et al., 2018; Prati et al., 2017; Zhu et al., 2017; Zong et al., 2019; Zou & Yue, 2017). With large samples of crash data, Bayesian networks were shown to be effective in identifying complex interrelationships among multiple crash attributes and crash outcomes (de Oña et al., 2013; Prati et al., 2017).

## 3 Data

Crash data from the NHTSA CRSS database was used in this study. The CRSS is a United States crash database with crash data extracted from nationally sampled crash reports (NHTSA, 2018). A total of about 150,000 crash observations were included in the 2016-2018 CRSS crash data, representing about 20 million police-reported crashes. The CRSS database organizes data into crash, vehicle, person, and event levels. The four levels of data can be linked through unique IDs for crashes, vehicles, and persons. In this paper, the crash, vehicle, and event level data files were used.

## 3.1 Subsetting

Since intersection two-vehicle crashes were the focus of this paper, a subset of crashes of the 2016-2018 CRSS data was obtained by applying the criteria listed in Table 1. The criteria ensured a data set consisting of intersection two-vehicle crashes involving only passenger vehicles in motion. Crashes that involved emergency vehicles, were alcohol-related or happened at locations with special configurations (e.g., work zone) were excluded, since the crash dynamics, sequences, and ODDs can be significantly different under the impact of those factors. We are concerned that the nature of those "special" crashes could lead to drastically different sequences of events and outcomes from the "common" crashes, and a separate analysis would be needed develop relevant scenarios for "special" crashes. In this paper, we focused on generating scenarios for the "common" crashes. By applying the criteria, we obtained a data set consisting of 39,850 observations, representing nearly 6 million crashes.

**Table 1 Subsetting criteria**

| Variable (Data Level) | Value | Description of Criterion |
|---|---|---|
| VE_TOTAL (Crash) | = 2 | Exactly two vehicles involved in crash. |
| VE_FORMS (Crash) | =2 | Exactly two vehicle-in-transport involved in crash. |
| PVH_INVL (Crash) | = 0 | No parked/working vehicles involved. |
| RELJCT2_IM (Crash) | = 2 or 3 | Crash happened at an intersection or was intersection-related. |
| WRK_ZONE (Crash) | = 0 | No work zone at crash location. |
| ALCHL_IM (Crash) | $\neq 1$ | No alcohol-related crash. |
| BDYTYP_IM (Vehicle) | < 50 | Only automobile, utility vehicles or light trucks* involved. |
| TOW_VEH (Vehicle) | = 0 | No vehicle trailing involved. |
| BUS_USE (Vehicle) | = 0 | No bus involved. |
| SPEC_USE (Vehicle) | = 0 | No special use vehicles involved. |
| EMER_USE (Vehicle) | = 0 | No emergency use vehicles involved. |

Note: * Light trucks with Gross Vehicle Weight Rating (GVWR) ≤ 10,000 LBS.

## 3.2 Numbering Crash Participants

Each intersection two-vehicle crash has two participating motor vehicles. The CRSS numbered the two vehicles as Vehicle 1 and Vehicle 2 "sequentially", without specifying the basis for that ordering(NHTSA, 2018). Before analyzing the data and developing crash scenarios, we renumbered the vehicles based on their initial positions and trajectories in crashes. By renumbering the participating vehicles, we ensured that scenarios generated using the crash data have consistent vehicle alignments. In simulation-based tests with a two-vehicle crash scenario, the automated driving system (ADS) being tested would be aligned as one of the two participating vehicles (Scanlon et al., 2021). The ADS would be first aligned as Vehicle 1, with Vehicle 2 as an adversarial agent, and then be aligned as Vehicle 2, with Vehicle 1 as an adversarial agent. In that case, we can fully make use of one scenario and test the ADS on both roles in a two-vehicle crash.

To renumber the participating vehicles, information about vehicles' initial positions and trajectories was obtained from the PC23 Crash Type Diagram of CRSS, as shown in Figure 1. For two-vehicle crashes, there are 5 high-level categories including "Same Trafficway Same Direction", "Same Trafficway Opposite Direction", "Change Trafficway Vehicle Turning", "Intersect Paths", and "Miscellaneous". The 5 high-level categories split into 10 configurations (denoted as using letters from "D" to "M"), which split again into more specific crash types (with participating vehicles position and trajectory numbered from "20" to "99"). Not all crash types appeared in the intersection two-vehicle crash sample. Crash types not included in the sample were marked with a red slash in Figure 1.

Rules of vehicle renumbering are also illustrated in Figure 1. Participants marked with a red dot were numbered as Vehicle 1, participants with a blue dot were numbered as Vehicle 2, and participants with a gray dot had the original numbering kept. Take the "68-69" (left turn meets through) combination in crash configuration "J" as an example, the vehicle with an initial position and trajectory "68" (left turn) is numbered as Vehicle 1, and the vehicle with an initial position and trajectory "69" (through) was numbered as Vehicle 2.

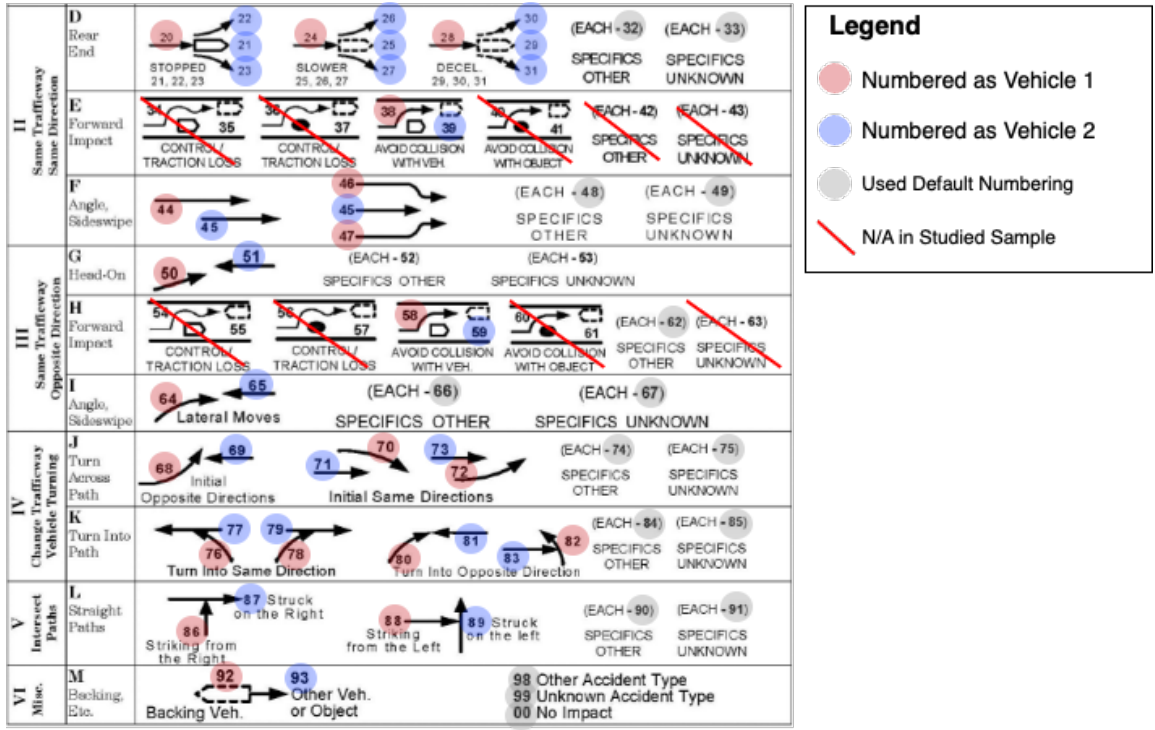

**Figure 1  CRSS two-vehicle crash types and participant renumbering (NHTSA, 2018)**

### 3.3    Encoding Sequences

The crash sequences were formed using four CRSS variables, PCRASH1 (pre-event movement), PCRASH2 (critical event pre-crash), and PCRASH3 (attempted avoidance maneuver) in the vehicle level data set VEHICLE, as well as the SOE (sequence of events) variable from the event level data set CEVENT. The PCRASH1~3 variables describe "what a vehicle was doing just prior to the critical precrash event", "what made the vehicle's situation critical", and "what was the corrective action made, if any, to this critical situation" (NHTSA, 2018). The SOE variable records series of harmful and non-harmful events occurred in the crashes, in chronological order.

The PCRASH1~3 and SOE events were combined following the rule illustrated in Figure 2. If the first event in SOE was a Vehicle 1 event, then the PCRASH1~3 events of Vehicle 1 were inserted before the PCRASH1~3 events of Vehicle 2, and all PCRASH events were inserted before SOE. Vice versa, if the first event in SOE was a Vehicle 2 event, then the PCRASH1~3 events of Vehicle 2 were inserted before the PCRASH1~3 events of Vehicle 1, and all PCRASH events were inserted before SOE.

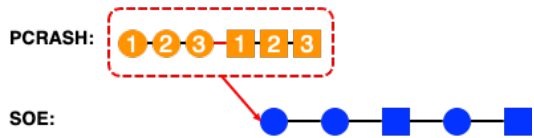
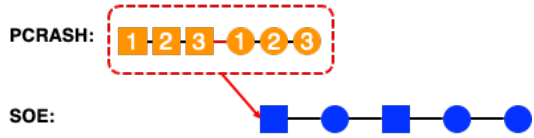
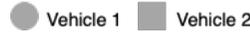

**Figure 2 Sequence structure**

Lengths (number of events) of the sequences ranged from 7 to 16, as shown in Table 2. With over 92% of the sequences having 7 events (with 6 PCRASH events and 1 SOE event), the average length was 7.2. Sequences longer than the average usually have more collision events. For example, the three sequences with 16 events were:

- 1L-1L-1N-2ST-2OEO-2N-2XV-1CARG-1ROR-1XF-1XF-1XF-2ROL-2XF-2XF-2XF
- 1ST-1ST-1N-2ST-2OET-2N-1XV-2ROL-2XF-2XF-1ROR-1XF-1XP-1XP-1XP-1XF
- 1L-1N-2ST-2OEO-2R-2XV-1XV-2ROR-1XF-1XF-1XO-1XO-1XO-1XO-1XO

Each of the three 16-event sequences describes a crash where run-off-the-road events happened after the collision between two vehicles, and multiple collisions with fixed object ("XF"), non-fixed objects ("XO"), or pedestrians ("XP") happened afterwards. For detailed event encoding information, please refer to Table A - 1 of Appendix A.

**Table 2 Sequence lengths**

|         | Sample size: 39,850 |
|---------|---------------------|
| Count   | Length              |
| 36,924  | 7                   |
| 844     | 8                   |
| 1327    | 9                   |
| 423     | 10                  |
| 194     | 11                  |
| 87      | 12                  |
| 41      | 13                  |
| 6       | 14                  |
| 1       | 15                  |
| 3       | 16                  |
| Average | 7.2                 |

CRSS has a total of 160 categories of pre-crash and collision events, including 20 in PCRASH1, 57 in PCRASH2, 14 in PCRASH3, and 69 in SOE. With two participating vehicles, the total number would be 320. With a length of 16, there would be theoretically $320^{16}$ possible sequences. Identifying patterns in sequences would be difficult with too many event categories. Therefore, in this study, the events were encoded by consolidating events that are similar in nature. With the encodings, the total number of event categories became 71, including 14 in PCRASH1, 29 in PCRASH2, 11 in PCRASH3, and 17 in SOE. Details of event encodings are in Table A - 1 of Appendix A.

### 3.4 Other Crash Attributes

To specify crash scenarios, variables describing crash outcomes, human factors, and environmental conditions were needed, in addition to the crash sequence patterns. The crash outcomes help identify scenarios with more severe injuries and fatalities, which may be of more interest in testing AVs. The human factors and environmental conditions help define ODDs which consist of other moving objects and static surroundings. Crash outcome variables used in this paper are summarized in Table 3. Variables describing human factors and environmental conditions are summarized in Table 4.

**Table 3  Crash outcomes**

| Variable | % | Variable | % |
|---|---|---|---|
| **Maximum severity (maxsev)** | | **Manner of collision (moc)** | |
| No apparent injury | 52.5% | Angle | 47.0% |
| Possible injury | 27.7% | Front-to-rear | 38.7% |
| Suspected minor injury | 11.5% | Sideswipe, same direction | 7.0% |
| Suspected serious injury | 7.2% | Front-to-front | 5.3% |
| Fatal | 0.4% | Sideswipe, opposite direction | 1.0% |
| Injured, severity unknown | 0.5% | (Other) | 1.0% |

Note: Sample size 39,850.

**Table 4  Human factors and environmental conditions**

| Variable | % | Variable | % | Variable | % |
|---|---|---|---|---|---|
| **Driver speeding (speeding)** | | **Urbanicity (urbrur)** | | **Speed limits (spdlim)** | |
| N+N | 90.5% | Urban | 80.2% | 45+45 | 20.7% |
| Y+N | 5.7% | Rural | 19.8% | 35+35 | 18.0% |
| U+N | 2.2% | **Time of day (tod)** | | Unknown | 13.4% |
| N+Y | 0.6% | Day | 80.7% | 40+40 | 10.4% |
| N+U | 0.5% | Night | 19.3% | 25+25 | 7.7% |
| U+U | 0.4% | **Lighting condition (light)** | | 30+30 | 6.2% |
| (Other) | 0.1% | Daylight | 77.1% | (Other) | 23.5% |
| **Careless driving (careless)** | | Dark-Not Lighted | 3.6% | **Road surface condition (surcon)** | |
| N+N | 90.3% | Dark-Lighted | 15.3% | Dry | 83.2% |
| N+Y | 1.0% | Dawn | 1.2% | Wet | 13.6% |
| Y+N | 8.5% | Dusk | 2.5% | Unknown | 1.1% |
| Y+Y | 0.2% | Dark-Unknown Lighting | 0.3% | Snow | 0.9% |
| **Driver did not see (didnotsee)** | | (Other) | 0.0% | Ice/frost | 0.5% |
| N+N | 99.0% | **Weather (weather)** | | Non-trafficway or driveway access | 0.4% |
| N+Y | 0.1% | Clear | 73.2% | (Other) | 0.2% |
| Y+N | 0.8% | Cloudy | 15.9% | **Traffic control devices (tcd)** | |
| Y+Y | 0.1% | Rain | 9.1% | Signal+Signal | 49.1% |
| **Reckless driving (reckless)** | | Snow | 1.4% | No TCD+No TCD | 19.7% |
| N+N | 96.7% | Fog, Smog, Smoke | 0.2% | Sign+No TCD | 11.7% |
| N+Y | 0.3% | Sleet or Hail | 0.1% | Sign+Sign | 7.8% |
| Y+N | 3.0% | (Other) | 0.2% | No TCD+Sign | 5.1% |
| Y+Y | 0.0% | **Type of intersection (typint)** | | Unknown+Unknown | 2.3% |
| **Improper control (impropctrl)** | | (Other) | 0.7% | (Other) | 4.3% |
| N+N | 99.5% | 3-Legged | 22.0% | | |
| N+Y | 0.0% | Unknown | 21.8% | | |
| Y+N | 0.5% | 4-Legged | 55.6% | | |

Note: Sample size 39,850.
Labels with "+" indicate conditions of Vehicle 1 on the left side and Vehicle 2 on the right side.
In speeding, etc.: U = Unknown; Y = Yes; N = No.

More than half (52.5%) of the sampled crashes ended with a maximum injury severity of no apparent injury, less than half (46.9%) with injury to different extents, and 0.4% with fatality. In terms of manner of collision, angle crashes made up 47.0% of the sample crashes, front-to-rear crashes made up 38.7%, and others made up 14.3%. The crash outcomes reflect a motor vehicle's exposure to crash risks. When setting up test scenarios for AV safety evaluation, the exposure can be adjusted through selecting a combination of crash sequence types.

In this paper, the human factors variables were derived from the "D22 Speeding Related" and "D24 Related Factors – Driver Level" variables of the CRSS VEHICLE data file. We focused on the most frequent driver factors involved in the sample crashes, including speeding, careless driving, did not see, reckless driving, and improper control. In terms of environmental condition variables, we included urbanicity, time of day, lighting condition, weather, type of intersection, speed limits, road surface condition, and traffic control device. For test scenario specifications, the human factors and environmental conditions describe the ODDs.

## 4  Methodology

This paper employed a procedure of two steps to specify crash scenarios which include crash sequences and depictions of ODDs. The first step was characterization of crash sequences using sequence analysis, and the second step was ODD specification based on a Bayesian network modeling of relationships among crash sequences, outcomes, human factors, and environmental conditions. Sequences and ODDs make up scenarios. Crash sequences describe what could happen between moving objects in the scenarios. ODDs illustrate the surrounding environment in the scenarios.

A crash sequence describes the progression of events that happened in a crash over time. It naturally provides a structure to model complex interactions between vehicle, driver, and the surrounding environment. The reasons that a sequence clustering + Bayesian network design was implemented in this paper rather than a more complex modeling design such as a dynamic Bayesian network were: 1) Such a model is simple and easy to interpret. 2) For a descriptive scenario library, the model offers enough detailed information on the co-occurrence of certain types of crash sequences and ODD settings. 3) Historical crash data only allows us to model interactions to such a level of detail. The CRSS crash sequences themselves include some information about vehicles' interaction with elements of the ODD such as roadside objects. Relative to the interactions between moving objects which changes rapidly, the ODD variables added in the Bayesian network modeling can be considered static.

### 4.1  Crash Sequence Comparison and Clustering

In sequence analysis, the encoded crash sequences were compared and grouped. The basis of sequence comparison and clustering is the measure of dissimilarity, which quantifies the difference between each pair of sequences (Cornwell, 2015; Studer & Ritschard, 2014, 2016). Optimal matching (OM) based dissimilarity measures have been widely used in bioinformatic and sociological research to for gene sequence or life course sequence analysis (Abbott, 1983, 1995; Cornwell, 2015; Kruskal, 1983; Studer & Ritschard, 2014, 2016). Some other studies also found OM based dissimilarity measures to be appropriate for crash sequence analysis (Song, 2021; Song et al., 2021, 2022). In the same study, we proposed a methodology to select the optimal encoding schemes and dissimilarity measures for different crash analysis use cases and found the Levenshtein distance to be an overall good choice for measuring crash sequence dissimilarity.

The OM takes two sequences and aligns them. The alignment of sequences involves several operations such as substitutions, deletions/insertions (or indels), compression and expansions, and transpositions (or swaps) (13, 18, 30). The Levenshtein distance uses only substitutions and indels, with fixed and unified costs (e.g., substitution = indel = 1). The mathematical expression of OM based dissimilarity between a pair of sequences, $x$ and $y$, is:

$$d_{OM}(x,y) = \min_j \sum_{i=i}^{\ell_j} \gamma(T_i^j) \qquad [1]$$

where $\ell_j$ denotes the transformations needed to turn sequence x into y;  and

$\gamma(T_i^j)$ is the cost of each elementary transformation $T_i^j$ (e.g., indel or substitution).

An example of sequence alignment using Levenshtein distance is shown in Table 5. There are multiple ways to align two sequences, "ABCD" and "ACB". Using a substitution cost of s and an indel cost of d, the two ways of alignment shown in Table 5 yielded different total costs. The OM then applies a greedy algorithm to return the minimum alignment cost as the dissimilarity between the two sequences.

**Table 5  Sequence alignment**

| Sequence 1 | A | B | C | D | |
|---|---|---|---|---|---|
| Sequence 2 | A | C | B | | |
| **Alignment 1** | | | | | |
| Sequence 1 | A | | B | C | D |
| Sequence 2 | A | C/ | B | ø | ø |
| Cost | | d | | d | d | = 3d |
| **Alignment 2** | | | | | |
| Sequence 1 | A | B | C | D | |
| Sequence 2 | A | ø | C | B | |
| Cost | | d | | s | = d+s |

Note: Insertion is marked with ø,
        Deletion is marked with slash/,
        Substitution is marked with underline

The Needleman-Wunsch algorithm is a classic OM algorithm to align sequences and find sequence dissimilarities (Needleman & Wunsch, 1970). For two sequences, A and B, an empty matrix L, of size length(A)+1 by length(B)+1 is created. Based on a set of indel and substitution costs, the algorithm fills matrix L and returns the smallest alignment cost (i.e., the dissimilarity) between sequences A and B. Pseudocode of the Needleman-Wunsch algorithm is as follows (Song et al., 2021).

**Algorithm** Needleman-Wunsch(A, B)

```
# initialize
L <- matrix of size length(A)+1 * length(B)+1
d <- indel cost
s <- substitution cost
# fill the cells of L
for i = 0 to length(A)
        L(i,0) <- d*i
for j = 0 to length(B)
        L(0, j) <- d*j
for i = 1 to length(A)
        for j = 1 to length(B) {
                insert <- L(i, j−1) + d
                delete <- L(i−1, j) + d
                substitute <- L(i−1, j−1) + s
                L(i, j) <- max(insert, delete, substitute)
                }
# smallest alignment cost (distance)
return L(length(A), length(B))
```

Using the Levenshtein distance, a dissimilarity matrix for a set of crash sequences is calculated. The matrix is of size n*n , where n is the number of sequences in the set. Each element in the matrix indicates the dissimilarity between a pair of sequence. The dissimilarity matrix can then be used as an input for a clustering algorithm to characterize crash sequences as distinctive types.

For clustering, we employed a weighted k-medoid method in this paper. K-medoid clustering has been applied in prior studies for crash characterization due to its good performance with categorical data and robustness against outliers (Nitsche et al., 2017; Song et al., 2021). The weighted k-medoid clustering algorithm used in this paper was developed by Studer, accommodating sampling weights (provided by CRSS data sets) in the clustering (Studer, 2013). The sequence dissimilarity calculation and sequence clustering were completed using R and the libraries "TraMineR" and "WeightedCluster" (Gabadinho et al., 2011; R Core Team, 2019; Studer, 2013).

To characterize the intersection two-vehicle crashes, the sequence comparison and clustering were done under the existing CRSS crash configuration (CC) classification. The distribution of CC is shown in Table 6. Sequence comparison and clustering was done for each CC category. By conducting sequence analysis using existing CC classification, we kept the rarer crash types which would otherwise be overlooked if all 39,850 crash sequences were analyzed in a lump sum. Those rarer crash types are especially useful for developing potentially challenging test scenarios.

**Table 6  Distribution of CRSS intersection two-vehicle crash configurations**

| Category | CC | Description | Count | % | Weighted Count | % |
|---|---|---|---|---|---|---|
| Same Trafficway, Same Direction | D | Rear End | 14,784 | 37.10% | 2,300,603 | 39.08% |
| | E | Forward Impact | 8 | 0.02% | 1,143 | 0.02% |
| | F | Angle, Sideswipe | 1,692 | 4.25% | 301,033 | 5.11% |
| Same Trafficway, Opposite Direction | G | Head-On | 153 | 0.38% | 14,178 | 0.24% |
| | H | Forward Impact | 9 | 0.02% | 586 | 0.01% |
| | I | Angle, Sideswipe | 117 | 0.29% | 18,069 | 0.31% |
| Change Trafficway, Vehicle Turning | J | Turn Across Path | 7,932 | 19.90% | 1,095,637 | 18.61% |
| | K | Turn Into Path | 6,777 | 17.01% | 1,010,868 | 17.17% |
| Intersect Paths | L | Straight Paths | 7,160 | 17.97% | 938,248 | 15.94% |
| Miscellaneous | M | Backing, Etc. | 1,218 | 3.06% | 207,224 | 3.52% |
| | | *Total* | *39,850* | *100.00%* | *5,887,588* | *100.00%* |

## 4.2 Bayesian Network Modeling

Bayesian network modeling has been used in prior studies to evaluate factors affecting crash type and injury severity (Davis, 2001, 2003; Simoncic, 2004; Davis et al., 2011; Fenton & Neil, 2011; Zhu et al., 2017; Zou & Yue, 2017; Ma et al., 2018; de Oña et al., 2013; Prati et al., 2017). Bayesian networks are directed acyclic graphs (DAGs), with nodes denoting variables and directed edges denoting the dependencies between variables (Pearl, 1985). The strengths of influences between variables are measured by conditional probabilities. If the graph has variables $x_1, \ldots, x_n$, and $S_i$ as the set of parents of $x_i$, an estimated conditional probability is then $P'(x_i|S_i)$. The following joint probability distribution exists for the graph:

$$P(x_1, \ldots, x_n) = \prod_i P'(x_i|S_i) \qquad [2]$$

A Bayesian network can be specified based on expert opinions when the number of variables is small and variable relationships are intuitive and simple. Oftentimes, defining a network is too complicated for humans and would need a data-driven approach to accomplish (Koller & Friedman, 2009). The construction of a Bayesian network consists of two steps (Zhu et al., 2017; Zou & Yue, 2017):

- Structure learning: determine selection of variables (nodes) and determine the dependencies or independencies between nodes, to form a DAG.
- Parameter learning: based on the determined DAG, estimate a conditional probability table for each node to quantify relationship between nodes.

In this paper, the structure learning was completed using a hill climbing algorithm, which finds the network structure with the highest Akaike information criterion (AIC) score. The AIC score used in the R library, "bnlearn" is calculated as the classic definition rescaled by –2 (Scutari, 2010):

$$AIC = \ln(\hat{L}) - 2k \qquad [3]$$

where $\hat{L}$ = the estimated maximum likelihood of the model; k = the number of estimated parameters. Therefore, a higher AIC score means a better Bayesian network model. The learned network structure was slightly adjusted based on the authors' domain knowledge in traffic crashes. Multiple networks structures were generated and compared to determine a most appropriate one for ODD specification, based on the conditional probability table. The structure and parameter learning were completed using R and the "bnlearn" library (R Core Team, 2019; Scutari, 2010). Bayesian networks were visualized using the "Rgraphviz" library (Hansen et al., 2021).

Based on a determined Bayesian network, we could understand the direct and indirect dependencies among variables including sequence types (developed from the sequence analysis), crash outcomes (manner of collision and injury severity), human factors, and environmental conditions, using graph visualizations. We could also identify the sequence types likely yielding serious crash outcomes and the ODD settings for specific sequence types by querying for conditional probabilities in the network. Scenarios can be defined using combinations of sequence types and ODD settings, and can be used to help render simulation tests for AV safety evaluation.

## 5 Results

This section presents the results of 1) crash sequence characterization from sequence analysis and 2) variable dependencies from Bayesian network modeling. Finally, an example of scenario specification based on the crash sequence type and Bayesian network is provided at the end of this section.

### 5.1 Sequence Types

As mentioned, sequence clustering was carried out for each CC category using Levenshtein distance and weighted k-medoid clustering. To measure the quality of clustering and determine the appropriate number of clusters, clustering quality indices including the Weighted Average Silhouette Width (ASWw), Hubert's Gamma (HG), Point Biserial Correlation (PBC), and Hubert's C (HC) were calculated. We used k values ranging from 2 to 25 to cluster the sequences and plotted the indices for comparison.

The plots in Figure 3 illustrate the clustering quality indices for the clustering of CC D (Same Trafficway, Same Direction – Rear End, 39% of all crashes) crash sequences. An optimal k value would give us maximum ASWw, HG, and PBC (all range from -1 to 1), and a minimum HC (ranges from 0 to 1). Figure 3(a) shows the original values of the four indices but is difficult to read, since the indices have different average values. Figure 3(b) shows the standardized index curves for easier comparison, and it shows that when k = 12 the ASWw, HG, and PBC all reach satisfactory high levels and at the same time HC reaches a satisfactory low level. Therefore, k = 12 was chosen as the number of clusters for CC D sequence clustering. The same procedure of plotting quality indices and determining the appropriate k value was applied to all CC categories.

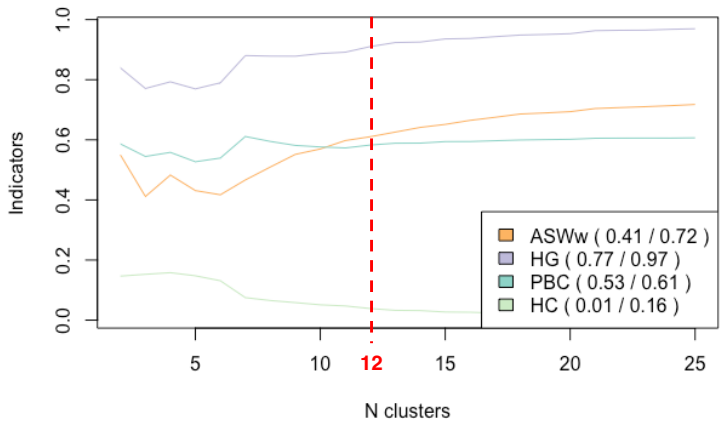

**(a) Original values**

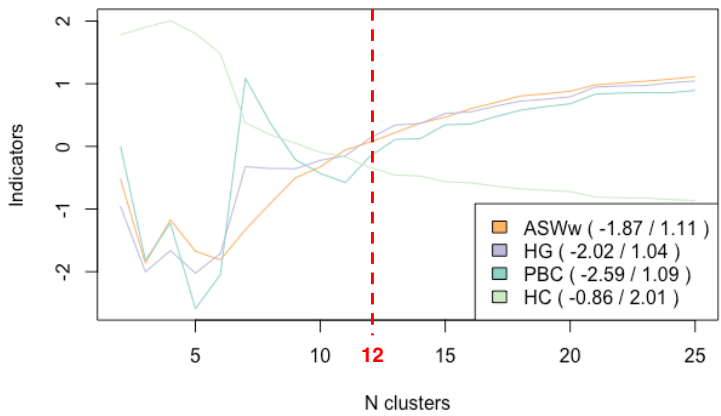

**(b) Standardized values**

**Figure 3  Clustering quality indices for CC D (rear end) sequences**

The clustering results are summarized in Table 7. In each cluster, only the sequence representing the most sequences is presented. More detailed results showing the top three representative sequences are shown in Table A - 2 of Appendix A. CC D Same Trafficway, Same Direction – Rear End (39% of all) crashes were characterized as 12 sequence types, CC J – Change Trafficway, Vehicle Turning – Turn Across Path (18% of all) crashes were characterized as 3 sequence types, CC K – Change Trafficway, Vehicle Turning – Turn Into Path (17% of all) crashes were characterized as 9 sequence types, and CC L – Intersect – Straight Paths (16% of all) crashes were characterized as 14 sequence types.

Interpretation of the representative sequences is presented in Table 7 for easier understanding of the sequence types. For example, the representative sequence of Type e3, in coded form, was "1ST-1OES-1BR-2ST-2OIS-2NA-1XV-1ROR-1XF-1NCH", and was interpreted as "v1 moving straight-other encroached into lane SD (brake and turned right) → v2 moving straight (no)", meaning Vehicle 1 and Vehicle 2 were both moving straight along the same direction. Some other vehicle encroached into Vehicle 1's lane, making Vehicle 1 brake and steer right. Vehicle 1 then collided into the rear of Vehicle 2, which did not make any maneuver to avoid the collision. Following the collision, Vehicle 1 ran off the road, hit a fixed object, and suffered a non-collision harmful event.

For the development of AV testing scenarios, it is important to balance details and abstraction, as rare but safety-critical crashes may be buried into larger groups of more common crashes in the clustering process.

By further characterizing intersection two-vehicle crashes under the existing crash configurations, some rare crash configurations such as E (Same Trafficway, Same Direction – Forward Impact) and G (Same Trafficway, Opposite Direction – Head On) could be represented and considered in scenario specification. Within each crash configuration category, the sequence clustering was able to identify different pattens in the pre-crash events and effectively assign sequences into distinct clusters. For example, the first three clusters under the D (Same Trafficway, Same Direction, Rear-End) crash configuration, as shown in Table 7, represent different initial states: d1 has Vehicle 1 moving straight and Vehicle 2 decelerating, d2 with Vehicle 1 making a right turn and Vehicle 2 stopped, and d3 with Vehicle 1 negotiating a curve and Vehicle 2 stopped. In addition to differences in initial states, types of collision avoidance maneuvers were also differentiated by the sequence clustering. Therefore, the resulting sequence types provide useful information for determining the initial locations, actions, and behavior of agents when recreating the scenarios in simulation.

Table 7  Sequence clustering results

| Category | Crash Configuration | Type | Interpretation of Representative Sequence | % |
|---|---|---|---|---|
| Same Trafficway, Same Direction | D-Rear End | d1 | v1 moving straight (decelerated) → v2 decelerating | 0.7 |
| | | d2 | v1 turning right → v2 stopped (no) | 0.8 |
| | | d3 | v1 negotiating curve → v2 stopped (no) | 1.4 |
| | | d4 | v1 moving straight → v2 decelerating | 5.4 |
| | | d5 | v1 accelerating → v2 stopped (no) | 2.0 |
| | | d6 | v2 stopped (no) → v1 moving straight (no) | 0.7 |
| | | d7 | v1 moving straight (decelerated) → v2 stopped (no) | 3.5 |
| | | d8 | v1 moving straight → v2 moving straight | 1.8 |
| | | d9 | v1 moving straight → v2 stopped | 1.2 |
| | | d10 | v1 decelerating → v2 stopped (no) | 1.5 |
| | | d11 | v1 moving straight (no) → v2 stopped (no) | 2.9 |
| | | d12 | v1 moving straight → v2 stopped (no) | 17.1 |
| | E-Forward Impact | e3 | v1 moving straight-other encroached into lane SD (brake and turned right) → v2 moving straight (no) | 0.0 |
| | F-Angle, Sideswipe | f1 | v2 moving straight-encroached into lane on left → v1 moving straight | 0.2 |
| | | f2 | v1 changing lane → v2 stopped (no) | 0.4 |
| | | f3 | v2 moving straight → v1 changing lane-encroached into lane on left | 0.3 |
| | | f4 | v1 changing lane-encroached into lane on left → v2 stopped (no) | 0.5 |
| | | f5 | v1 changing lane-encroached into lane on right (no) → v2 moving straight | 0.2 |
| | | f6 | v2 moving straight-encroached into lane on left → v1 stopped (no) | 0.2 |
| | | f7 | v1 moving straight-encroached into lane on right → v2 moving straight | 0.3 |
| | | f8 | v1 changing lane-encroached into lane on left → v2 moving straight | 1.2 |
| | | f9 | v1 changing lane-encroached into lane on right → v2 moving straight | 1.7 |
| Same Trafficway, Opposite Direction | G-Head-On | g1 | v1 moving straight-encroached into lane on left → v2 moving straight | 0.2 |
| | | g2 | v1 moving straight (steering left) → v2 stopped (no) | 0.0 |
| | H-Forward Impact | h1 | v1 moving straight (steering left) → v2 moving straight | 0.0 |
| | I-Angle, Sideswipe | i1 | v1 moving straight-encroached into lane on left → v2 stopped (no) | 0.2 |
| | | i2 | v2 moving straight-other encroached into lane → v1 moving straight-speeding | 0.1 |

Note: v1 = Vehicle 1; v2 = Vehicle 2. See Section 3.2 Numbering Crash Participants for details.
    SD = same direction.
    The arrow symbol, "→", means that the vehicle left of the arrow collided into the vehicle right of it.
    Content in the parentheses is crash avoidance maneuver, no parenthesis means action unknown.

(Table 7 Continued)

| Category | Crash Configuration | Type | Interpretation of Representative Sequence | % |
|---|---|---|---|---|
| Change Trafficway, Vehicle Turning | J-Turn Across Path | j1 | v1 turning right → v2 moving straight-other encroached into lane SD | 1.5 |
| | | j2 | v1 turning left → v2 moving straight-other encroached into lane OD | 9.2 |
| | | j3 | v2 moving straight-other encroached into lane OD → v1 turning left | 7.8 |
| | K-Turn Into Path | k1 | v1 turning left → v2 stopped-other turning into OD (no) | 0.9 |
| | | k2 | v2 moving straight → v1 turning left-other encroached on cross street | 1.4 |
| | | k3 | v2 moving straight-other encroached on OD → v1 turning left | 3.9 |
| | | k4 | v1 turning left → v2 moving straight-other encroached into lane OD | 3.1 |
| | | k5 | v1 turning left → v2 stopped-other turning into OD (no) | 0.4 |
| | | k6 | v1 turning right → v2 moving straight-other encroached into lane SD | 3.0 |
| | | k7 | v2 moving straight-other turning into lane SD → v1 turning right | 0.9 |
| | | k8 | v1 turning left-other encroached into lane from cross street → v2 moving straight | 0.5 |
| | | k9 | v1 turning left → v2 moving straight-other encroached into lane SD | 3.2 |
| Intersect Paths | L-Straight Paths | l1 | v2 moving straight-other encroached into lane CS → v1 moving straight | 0.2 |
| | | l2 | v1 accelerating/start moving → v2 moving straight-other encroached into lane CS | 0.5 |
| | | l3 | v1 moving straight-other encroached into lane CS → v2 accelerating/start moving | 0.4 |
| | | l4 | v1 moving straight-other encroached into lane CS → v2 moving straight-other encroached into lane CS | 0.3 |
| | | l5 | v1 moving straight-other encroached into lane CS (brake) → v2 moving straight | 0.2 |
| | | l6 | v2 moving straight → v1 moving straight-other encroached into lane CS | 1.5 |
| | | l7 | v1 moving straight → v2 moving straight-other encroached into lane CS | 0.2 |
| | | l8 | v1 moving straight (no) → v2 moving straight-other encroached into lane CS (no) | 0.2 |
| | | l9 | v1 moving straight (no) → v2 moving straight-other encroached into lane CS | 0.7 |
| | | l10 | v1 moving straight-other encroached into lane → v2 moving straight (no) | 0.6 |
| | | l11 | v1 moving straight → v2 moving straight | 0.7 |
| | | l12 | v1 moving straight-other encroached into lane CS → v2 moving straight | 3.6 |
| | | l13 | v1 moving straight → v2 moving straight-other encroached into lane CS | 6.6 |
| | | l14 | v1 moving straight (brake) → v2 moving straight-other encroached into lane CS | 0.3 |
| Miscellaneous | M-Backing, Etc. | m1 | v1 backing up → v2 stopped-other backing (no) | 1.5 |
| | | u | v1 making U-turn → v2 moving straight-other encroached into lane OD | 2.0 |

Note: SD = same direction; OD = opposite direction; CS = cross street.

## 5.2 Variable Dependencies

Using a hill climbing algorithm with AIC as the criterion for model selection, a Bayesian network was learned to illustrate the relationships among sequence types, crash outcomes, human factors, and environmental condition variables. The network is shown in Figure 4. Each node represents a variable, and the directed arcs represent dependencies among variables. The weight of an arc represents its strength, measured by the potential change in AIC score led by removal of the arc from the network (i.e., the difference between the network's AIC score with and without the arc) (Scutari, 2010). If the change in AIC is negative, that means removing the arc harmed the network by losing information. Therefore, a more negative difference indicates a higher arc strength (i.e., stronger relationship between two variables). The network was only slightly adjusted by removing an arc from lighting condition ("light") to "weather" to simplify the network without compromising the overall model AIC score.

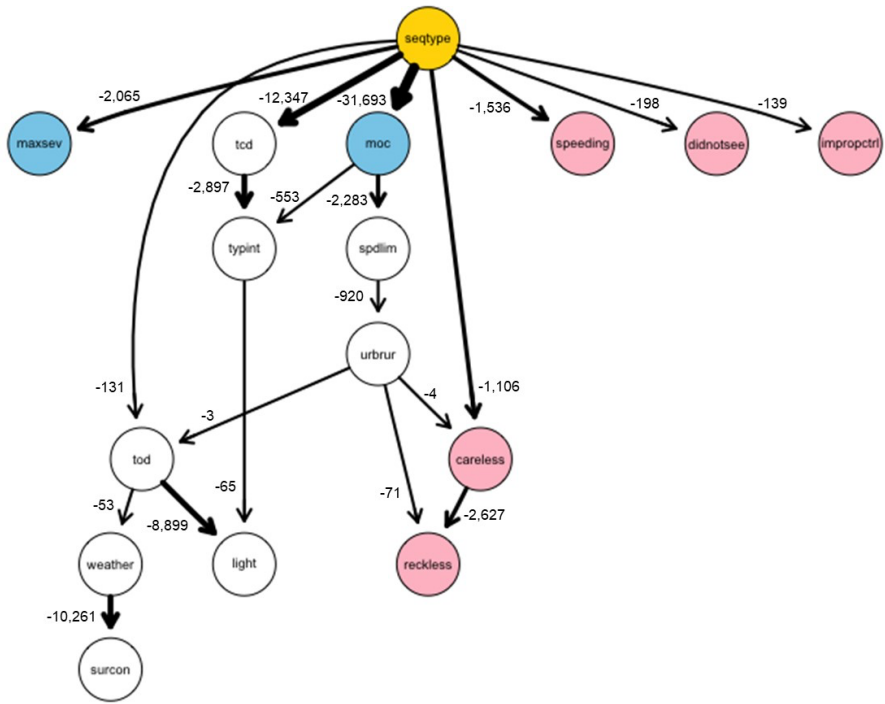

Note: Crash outcomes are in blue; human factors are in pink; environmental conditions are in white.
Numbers indicate strengths (change in AIC) of arc, the more negative, the stronger the link.

**Figure 4  Bayesian network generated from hill climbing learning**

The Bayesian network shows that sequence type had strong relationships with crash outcome variables – maximum injury severity ("maxsev") and manner of collision ("moc"). Sequence type was also directly or indirectly associated with human factors and environmental conditions. Five strongest direct links with sequence type were manner of collision ("moc"), traffic control device ("tcd"), maximum injury severity ("maxsev"), speeding, and careless driving ("careless").

The variable pointed by the arrowhead of an arc is directly dependent on the variable on the other end of the arc. Note that the arcs pointing from crash sequence type to crash outcomes indicated expected dependencies, as supported by prior studies' findings that the order of events and actions taken during the pre-crash and crash periods directly affect manner of collision and injury severity (Wu et al., 2016). As mentioned, a Bayesian network can be specified by a human expert or be constructed using data-driven approaches.

Therefore, depending on how much influence of expert opinions was incorporated in the specification of Bayesian networks, prior studies suggested different network structures regarding the relationships between crash outcomes and human and environmental factors (Ma et al., 2018; Zou & Yue, 2017; de Oña et al., 2013; Prati et al., 2017). In the studies analyzing crash causations, expert opinions heavily affected the Bayesian network design, thus arcs pointed from human and environmental factors to crash outcomes (Ma et al., 2018; Zou & Yue, 2017). However, the studies following a data-driven approach showed that Bayesian networks learned from crash data had some arcs pointing from crash outcomes to human and environmental factors (de Oña et al., 2013; Prati et al., 2017). In this study, the Bayesian network learned from crash data, as shown in Figure 4, has similar arc patterns with the latter studies applying a primarily data-driven approach.

To confirm that the primarily data-driven approach generated a better network than an expert-opinion oriented approach, an alternative Bayesian network was constructed with more influence of the authors' intuitions and domain knowledge, as shown in Figure 5. In the alternative network, arcs were manually added to point from human factors and environmental conditions to sequence types. The resulted network shows very weak arc strengths. Therefore, we followed the data-driven approach and selected the Figure 4 network as the final Bayesian network for test scenario specification.

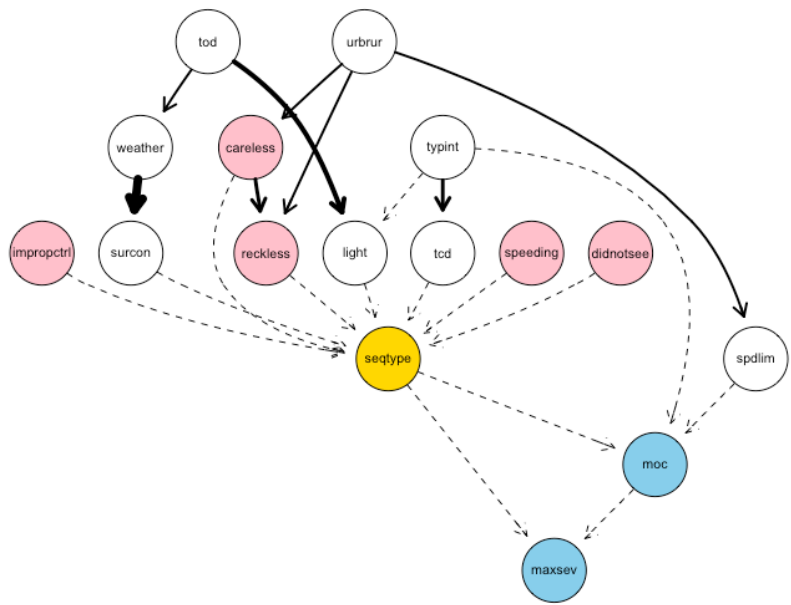

Note: Dashed arcs indicate weak strengths (> 0).

**Figure 5 Alternative Bayesian network**

A note on interpreting the inter-variable relationships in the Bayesian network such as that between sequence type ("seqtype") and variables such as traffic control devices ("tcd") in the Figure 4 network: Based on intuition, one may think that sequence type should be dependent on traffic control device and not the other way around, but such an intuition-based dependency requires strong assumptions of the process of how a traffic control device would affect the formation of a certain sequence type. There is no variable or data available in the crash database to support such assumptions. On the other hand, the current dependency should be interpreted as "given the information of a sequence type, one would know the type of traffic control device that was installed at intersections where such type of sequence most likely occurred", which is strong and supported by the data.

To confirm the relationships between sequence types and crash outcomes, as well as between sequence types, human factors, and environmental conditions, we tested the local network structural stability by developing partial Bayesian networks as shown in Figure 6 and Figure 7. Figure 6 shows that the relationships between sequence types and crash outcomes did not change after removing all other variables from the original network (in Figure 4). Figure 7 shows that the local network structure changed only slightly on type of intersection ("typint") and speed limit ("spdlim") after crash outcome variables were removed from the original network. Therefore, the local network structures were stable, confirming the relationships between sequence types and other variables.

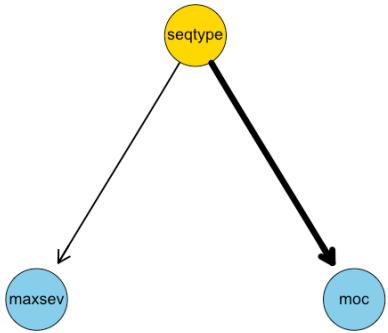

**Figure 6  Bayesian network of sequence types and crash outcomes**

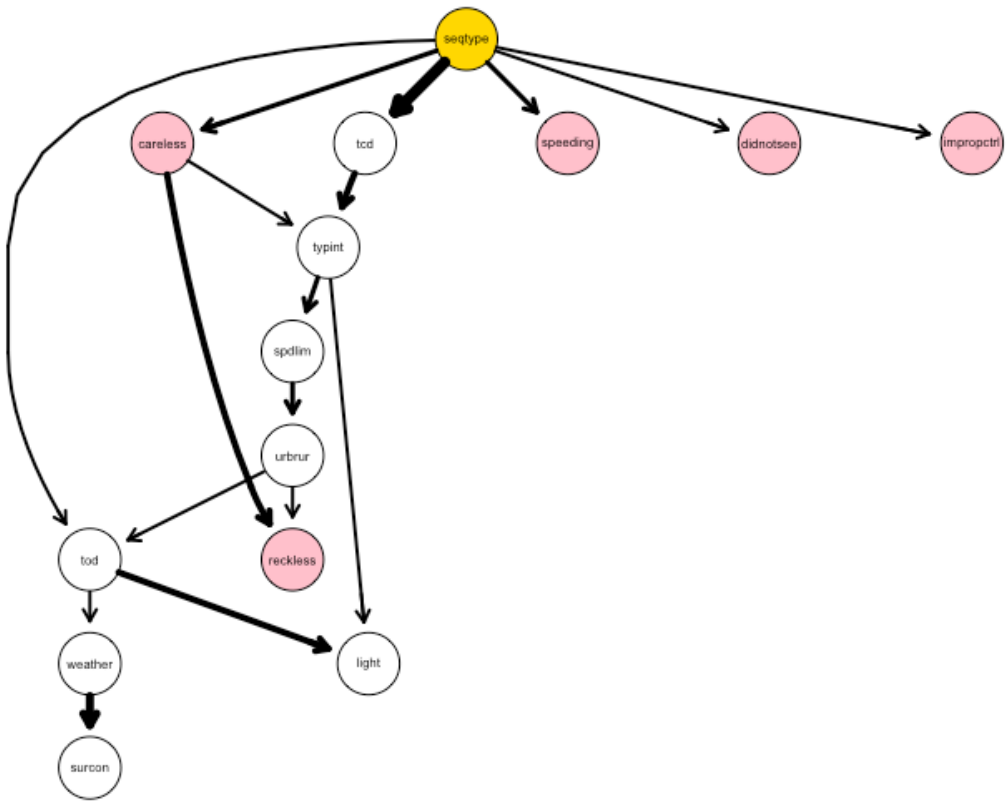

**Figure 7  Bayesian network of sequence types, human factors, and environmental conditions**

## 5.3 Scenario Specification

The final Bayesian network in Figure 4 is of two uses in specifying test scenarios. First, crash sequences of certain injury severity levels can be selected based on the dependencies between sequence types and crash outcomes. Second, after sequence types of interest are determined, their associated ODD attributes can be specified based on the dependencies among sequence types, human factors, and environmental conditions. The scenario specification can be done by querying the conditional probability table generated from the final Bayesian network. We demonstrate the process here with an example.

If we would like to specify scenarios for some intersection two-vehicle crashes that resulted in fatalities, we can first query the final Bayesian network to obtain the distribution of sequence types that resulted in fatalities, as shown in Figure 8. Using the "bnlearn" library in R, the distribution was generated based on Monte Carlo particle filters (Scutari, 2010). The query was run 1,000 times (1,000 replications), each time obtaining the counts of crashes under all sequence types. Then the average count was calculated within each sequence type over the 1,000 replications. The Figure 8 illustration presents the distribution of average counts. The distribution shows that sequence types j2, j3, k3, l7, l12, and l13 were the most frequent fatal crash sequences.

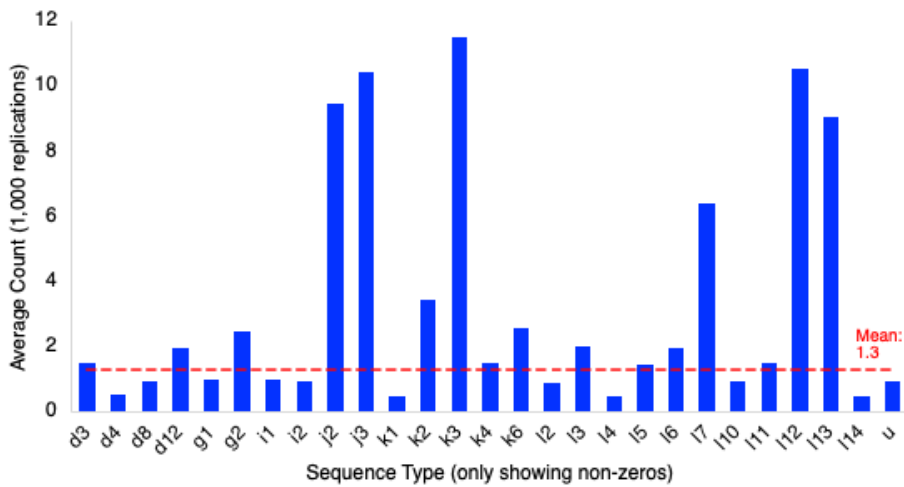

**Figure 8 Distribution of sequence types resulting in fatalities**

If we would like to specify ODDs for a sequence type, for example, k3, we can query the Bayesian network again for distributions of human factor and environmental condition variables using "seqtype = k3" as a criterion. The crash type k3 is illustrated in Figure 9, and it is a type of "changing traffic way turning into path" crash, where Vehicle 2 (V2 in figure) is moving straight and hitting Vehicle 1 (V1 in figure) which is turning left onto the trafficway that Vehicle 2 is on.

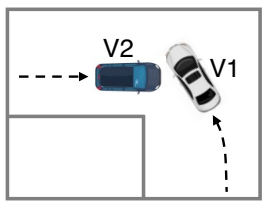

**Figure 9 Illustration of crash type k3**

Two examples of Bayesian network query for crash type k3 are demonstrated here. Table 8 shows query results for the distribution of intersection type and traffic control device (TCD) in k3 crashes. Table 9 shows query results for the distribution of drivers' speeding behavior and time of day in k3 crashes. The two tables are color coded to show larger values in darker green and smaller values in lighter green. From the query results we found that k3 crashes happened most frequently at 4-legged intersections with sign control on minor approaches, 3-legged intersections with sign control on minor approaches, and at 4-legged intersections with signal control on all approaches. Speeding was not related to 95% of k3 crashes. More k3 crashes happened in daytime than in nighttime, but the proportion of speeding-related crashes in all k3 crashes were the same 2% regardless of daytime or nighttime.

**Table 8  Distribution of intersection type and TCD in k3 crashes**

|  | Type of Intersection | | | |
|---|---|---|---|---|
| TCD | 4-Legged | 3-Legged | Roundabout | Other |
| No TCD+No TCD | 40.5 | 45.9 | 0.6 | 21.8 |
| No TCD+Sign | 20.0 | 4.0 | 0.1 | 1.6 |
| No TCD+Signal | 1.8 | 0.3 | 0.0 | 0.3 |
| Other+No TCD | 0.2 | 0.3 | 0.0 | 0.0 |
| Other+Other | 0.5 | 0.6 | 0.0 | 0.4 |
| Sign+No TCD | 217.4 | 184.9 | 4.8 | 30.0 |
| Sign+Other | 3.3 | 5.3 | 0.1 | 1.4 |
| Sign+Sign | 29.6 | 5.8 | 0.8 | 7.9 |
| Sign+Signal | 1.9 | 0.7 | 0.0 | 0.3 |
| Sign+Unknown | 8.9 | 7.1 | 0.0 | 9.2 |
| Signal+No TCD | 3.0 | 0.5 | 0.0 | 0.4 |
| Signal+Sign | 0.4 | 0.1 | 0.0 | 0.0 |
| Signal+Signal | 128.2 | 15.7 | 0.8 | 22.2 |
| Unknown+No TCD | 1.8 | 4.2 | 0.1 | 3.8 |
| Unknown+Sign | 2.7 | 0.6 | 0.0 | 2.4 |
| Unknown+Signal | 1.6 | 0.4 | 0.0 | 0.5 |
| Unknown+Unknown | 2.8 | 1.1 | 0.1 | 5.1 |

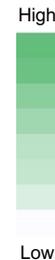

High

Low

Note: Average values of 1,000 replications.
Labels with "+" indicate conditions of Vehicle 1 on the left side and Vehicle 2 on the right side.

**Table 9  Distribution of speeding behavior and time of day in k3 crashes**

|  | Speeding | | | | | |
|---|---|---|---|---|---|---|
| Time of Day | N+N | N+U | U+N | U+U | N+Y | Y+N |
| Day | 671.4 | 6.6 | 7.6 | 5.3 | 9.6 | 2.8 |
| Night | 146.4 | 1.4 | 1.6 | 1.2 | 2.0 | 0.7 |

Note: Average values of 1,000 replications.
N = not speeding, U = unknown, Y = speeding.

With the information obtained from Table 8 and Table 9, we can specify several ODDs for the testing of k3 crash sequence type. For example, "a 4-legged intersection with sign control on the minor approaches at daytime". More complex queries can also be run to obtain more comprehensive descriptions of ODDs. Given an ODD, we can also obtain the distribution of sequence types by querying the Bayesian network and carry out tests accordingly. For example, Table 10 shows the query results of sequence type distribution at signal-controlled intersections. The results are color-coded to show the within-category (4-legged or 3 legged) distribution, with darker color indicating a higher proportion and lighter color indicating a lower proportion within category. Most frequently occurred crash sequence types for both 4-legged and 3-legged signalized intersections are i1, d12, j2, and j3. Crashes can be sampled based on this distribution and used to reconstruct scenarios in a simulation environment for AV testing.

**Table 10  Distribution of sequence types at signal-controlled intersections**

| Sequence Type | Type of Intersection | | Sequence Type | Type of Intersection | |
|---|---|---|---|---|---|
| | 4-Legged | 3-Legged | | 4-Legged | 3-Legged |
| d1 | 35.7 | 6.9 | j1 | 93.1 | 10.7 |
| d2 | 35.8 | 6.9 | j2 | 958.6 | 115.6 |
| d3 | 45.7 | 8.8 | j3 | 816.9 | 98.9 |
| d4 | 257.8 | 49.7 | k1 | 29.9 | 3.9 |
| d5 | 162.8 | 31.6 | k2 | 194.1 | 23.4 |
| d6 | 43.1 | 8.1 | k3 | 128.2 | 15.5 |
| d7 | 237.7 | 45.3 | k4 | 48.3 | 6.0 |
| d8 | 104.5 | 20.2 | k5 | 9.6 | 1.3 |
| d9 | 72.2 | 13.9 | k6 | 119.5 | 14.5 |
| d10 | 98.7 | 18.8 | k7 | 33.7 | 4.3 |
| d11 | 194.5 | 37.3 | k8 | 61.0 | 7.4 |
| d12 | 1171.2 | 225.9 | k9 | 89.2 | 10.9 |
| e3 | 1.1 | 0.2 | l1 | 13.5 | 1.7 |
| f1 | 14.2 | 1.6 | l2 | 7.3 | 0.9 |
| f2 | 23.3 | 2.7 | l3 | 3.5 | 0.4 |
| f3 | 14.8 | 1.7 | l4 | 23.2 | 2.8 |
| f4 | 39.6 | 4.4 | l5 | 8.9 | 1.0 |
| f5 | 7.0 | 0.8 | l6 | 90.3 | 11.0 |
| f6 | 9.7 | 1.1 | l7 | 14.5 | 1.7 |
| f7 | 16.9 | 1.8 | l8 | 10.5 | 1.3 |
| f8 | 57.0 | 6.1 | l9 | 25.2 | 3.0 |
| f9 | 90.9 | 9.8 | l10 | 15.8 | 1.9 |
| g1 | 17.5 | 2.1 | l11 | 98.5 | 12.0 |
| g2 | 13.2 | 1.9 | l12 | 226.1 | 27.7 |
| h1 | 1.8 | 0.2 | l13 | 453.4 | 55.5 |
| i1 | 2176.0 | 263.3 | l14 | 22.2 | 2.7 |
| i2 | 129.1 | 15.8 | m1 | 54.9 | 10.4 |
| | | | u | 98.0 | 12.7 |

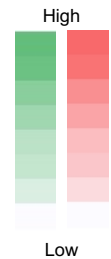

High

Low

Note: Average values of 1,000 replications.

Specified scenarios can be validated in simulation for their capabilities in testing AV safety. Based on the scenario design, moving agents (including an AV agent and agents simulating other road users) can be deployed and the ODD settings can also be rendered in the simulation. If the AV agent fails to maintain safety throughout the scenario simulation. The scenarios are then proved to be useful in capturing safety flaws of the AV agent. Various types of AV agents developed by different organizations can be included in the simulation testing to validate the reliability of the scenario library.

# 6 Discussion and Conclusion

This paper presents a procedure to generate crash scenarios for AV safety testing. The method consists of two steps, 1) characterization of crashes encoded by sequences of events using sequence analysis techniques, and 2) specification of scenarios based on a Bayesian network modeling dependencies among crash sequence types, crash outcomes, and variables depicting ODDs. The procedure was demonstrated using 2016-2018 intersection two-vehicle crash data from the NHTSA CRSS database.

This paper has two major findings. First, we characterized the intersection two-vehicle crashes as 55 types based on their patterns in sequences of events. The 55 crash sequence types offered more information about crash progression to the original CRSS configuration and helped identify rare crash types which could otherwise be overlooked. Second, we found the dependencies among crash sequence types, crash outcomes, human factors, and environmental conditions using a Bayesian network. Sequence types were found to be the core of the network and have direct effects on crash outcomes, as expected. The dependencies of human factors and environmental conditions on sequence types are useful in specifying ODDs for crash sequence types and identifying distributions of crash sequence types for specific ODDs.

The contribution of this paper is that it offers a methodology to systematically generate a crash scenario library based on national-level crash databases. Such a library would offer a comprehensive set of crash sequence types and ODDs, making it an appropriate guide for developing simulation-based tests for AV safety evaluation. Therefore, such a scenario library would complement scenario generating methodologies developed based on vehicle kinematic data sources such as naturalistic driving data and crash reconstruction data. Comparing with previous efforts in systematically generating scenarios for AV testing using crash data, such as the work by Nitsche et al. (2017), Sander and Lubbe (Sander & Lubbe, 2018), and the Safety Pool scenario library of Warwick University (Esenturk et al., 2021, 2022), the approach we proposed in this paper is a significant improvement. That is because 1) the crash sequence modeling captures the interactive dynamics and 2) the Bayesian network provides a comprehensive but interpretable representation of the relationships between all types of factors involved in crashes. Importantly, for specifying AV testing scenarios, the inclusion of crash dynamics conforms to the definition of "scenario" (Ulbrich et al., 2015); the model interpretability ensures that testers or experts are able to comprehend the meaning of each scenario; and the model comprehensiveness offers a large scenario space which allows the flexibility to develop tests targeting different levels of safety-criticality and various ODD settings.

Future work along this paper's line of research would bring improvements by addressing the limitations in crash sequence data source. More detailed crash event data than what we obtained from the CRSS database (or any other current national or state-level crash database in the United States) were not available, so crash sequences used in the case study were limited to the level of details provided by CRSS. More ODD attributes would benefit the Bayesian network modeling by providing more information. However, a more complex network would take more effort to interpret. For future work, data sources with more detailed crash sequence data would be sought and used to generate more detailed crash characterization. More elaborative Bayesian network models, such as dynamic Bayesian network, would be explored for use in modeling crash sequences and ODDs for AV test scenario specification. Also, an important next step would be the integration of crash sequences and vehicle kinematic information to define concrete scenarios that can be implemented in simulation-based testing platforms, as well as the validation of scenarios in testing AV safety performance.

**Funding**

This research was partially sponsored by the Safety Research using Simulation University Transportation Center (SAFER-SIM). SAFER-SIM is funded by a grant from the U.S. Department of Transportation's University Transportation Centers Program (69A3551747131). The work presented in this paper remains the responsibility of the authors.

# Appendix A
# Table A - 1  Event Encodings

**PCRASH1**

| Veh1 | Veh2 | CRSS Original Categories and Description |
|---|---|---|
| 1A | 2A | 3 Accelerating in Road, 4 Starting in Road |
| 1B | 2B | 2 Decelerating in Road |
| 1BU | 2BU | 13 Backing Up (Other Than for Parking Position) |
| 1C | 2C | 14 Negotiating a Curve |
| 1CA | 2CA | 17 Successful Corrective Action to a Previous Critical Event |
| 1E | 2E | 8 Leaving a Parking Position; 15 Changing Lanes; 16 Merging |
| 1L | 2L | 11 Turning Left |
| 1N | 2N | 0 No Driver Present/Unknown if Driver Present; 98 Other; 99 Unknown |
| 1P | 2P | 9 Entering a Parking Position |
| 1PA | 2PA | 6 Passing or Overtaking Another Vehicle |
| 1R | 2R | 10 Turning Right |
| 1S | 2S | 5 Stopped in Roadway; 7 Disabled or Parked in Travel Lane |
| 1ST | 2ST | 1 Going Straight |
| 1U | 2U | 12 Making a U-turn |

**PCRASH2**

| Veh1 | Veh2 | CRSS Original Categories and Description |
|---|---|---|
| | | *This Vehicle* |
| 1B | 2B | 18 This Vehicle Decelerating |
| 1BU | 2BU | 20 Backing |
| 1ELE | 2ELE | 12 Off The Edge of The Road on The Left Side |
| 1ELL | 2ELL | 10 Over The Lane Line on Left Side of Travel Lane |
| 1ERE | 2ERE | 13 Off The Edge of The Road on The Right Side |
| 1ERL | 2ERL | 11 Over The Lane Line on Right Side of Travel Lane |
| 1L | 2L | 15 Turning Left |
| 1LCF | 2LCF | 6 Traveling Too Fast For Conditions |
| 1LCM | 2LCM | 2 Stalled Engine; 4 Non-Disabling Vehicle Problem (e.g., Hood Flew Up); 5 Poor Road Conditions (Puddle, Pothole, Ice, etc.) |
| 1LCO | 2LCO | 8 Other Cause of Control Loss, 9 Unknown Cause of Control Loss |
| 1LCS | 2LCS | 1 Blow Out/Flat Tire, 3 Disabling Vehicle Failure (e.g., Wheel Fell Off) |
| 1N | 2N | 19 Unknown Travel Direction; 98 Other Critical Precrash Event; 99 Unknown |
| 1R | 2R | 16 Turning Right |
| 1ST | 2ST | 17 Crossing Over (Passing Through) Junction |
| 1U | 2U | 21 Making a U-Turn |
| | | *Other Vehicle* |
| 1OEN | 2OEN | Encroached Unknown Direction: 73 From Driveway, Intended Path Not Known; 78 Encroaching By Other Vehicle – Details Unknown |
| 1OEO | 2OEO | Encroached Opposite Direction: 62 From Opposite Direction Over Left Lane Line; 63 From Opposite Direction Over Right Lane Line; 67 From Crossing Street, Turning Into Opposite Direction; 72 From Driveway, Turning Into Opposite Direction |
| 1OES | 2OES | Encroached Same Direction: 60 From Adjacent Lane (Same Direction)-Over Left Lane Line; 61 From Adjacent Lane (Same Direction)-Over Right Lane Line; 64 From Parking Lane/Shoulder, Median/Crossover, Roadside; 65 From Crossing Street, Turning Into Same Direction; 70 From Driveway, Turning Into Same Direction; 74 From Entrance to Limited Access Highway |
| 1OET | 2OET | Encroached Across Path: 66 From Crossing Street, Across Path; 71 From Driveway, Across Path |
| 1OIN | 2OIN | In Lane: 59 Unknown Travel Direction Of The Other Motor Vehicle in Lane |
| 1OIO | 2OIO | In Lane: 54 Traveling in Opposite Direction |
| 1OIR | 2OIR | In Lane: 56 Backing |
| 1OIS | 2OIS | In Lane Same Direction: 50 Other Vehicle Stopped; 51 Traveling in Same Direction with Lower Steady Speed; 52 Traveling in Same Direction while Decelerating; 53 Traveling in Same Direction with Higher Speed |
| 1OIT | 2OIT | In Lane: 55 In Crossover |
| | | *Animal/Object/Bike/Ped* |
| 1AIA | | 88 Animal Approaching Road |
| 1AII | 2AII | 87 Animal in Road |
| | 2BII | 83 Pedalcyclist/Other Non-Motorist in Road |
| | 2OBI | 90 Object in Road |
| 1PII | 2PII | 80 Pedestrian in Road |

**PCRASH3**

| Veh1 | Veh2 | CRSS Original Categories and Description |
|---|---|---|
| 1A | 2A | 10 Accelerated |
| 1AL | 2AL | 11 Accelerating And Steering Left |
| 1AR | 2AR | 12 Accelerating And Steering Right |
| 1B | 2B | 15 Braking and Unknown Steering Direction; 16 Braking |
| 1BL | 2BL | 8 Braking And Steering Left |
| 1BR | 2BR | 9 Braking And Steering Right |
| 1L | 2L | 6 Steering Left |
| 1N | 2N | 0 No Driver Present/Unknown if Driver Present; 98 Other Actions; 99 Unknown/Not Reported |
| 1NA | 2NA | 1 No Avoidance Maneuver |
| 1R | 2R | 7 Steering Right |
| 1RB | 2RB | 5 Releasing Brakes |

**SOE**

| Veh1 | Veh2 | CRSS Original Categories and Description |
|---|---|---|
| 1AIR | 2AIR | 67 Vehicle Went Airborne |
| 1CARG | 2CARG | 60 Cargo/Equipment Loss or Shift (non-harmful) |
| 1CM | 2CM | 65 Cross Median; 68 Cross Centerline |
| 1ED | 2ED | 71 End Departure |
| 1EF | 2EF | 61 Equipment Failure (blown tire, brake failure, etc.); 62 Separation of Units |
| 1NCH | 2NCH | Non-collision Harmful Events: 2 Fire/Explosion; 3 Immersion or Partial Immersion; 4 Gas Inhalation; 5 Fell/Jumped from Vehicle; 6 Injured in Vehicle (Non-Collision); 7 Other Noncollision; 16 Thrown or Falling Object; 44 Pavement Surface Irregularity (Ruts, Potholes, Grates, etc.); 51 Jackknife (Harmful to This Vehicle); 72 Cargo/Equipment Loss, Shift, or Damage (Harmful) |
| 1RE | 2RE | 69 Re-entering Roadway |
| 1RLO | 2RLO | 1 Rollover/Overturn |
| 1RO | 2RO | 79 Ran off Roadway - Direction Unknown |
| 1ROL | 2ROL | 64 Ran Off Roadway-Left |
| 1ROR | 2ROR | 63 Ran Off Roadway-Right |
| 1XB | 2XB | Collision: With Pedalcyclist |
| 1XF | 2XF | Collision: With Fixed Object |
| 1XO | 2XO | Collision: With Other Object Not Fixed |
| 1XP | 2XP | Collision: With Pedestrian, Non-Motorist on Personal Conveyance |
|  | 2XPV | Collision: With Parked Motor Vehicle |
| 1XV | 2XV | Collision: With the Other Motor Vehicle In-Transport |

**Table A - 2  Intersection two-vehicle crash sequence types**

| Type | Weighted Count | % in Total | Representative Sequences | % in Type |
|---|---|---|---|---|
| d1 | 42,092 | 0.7% | 1ST-1OIS-1B-2B-2OIS-2N-1XV | 45% |
|  |  |  | 1ST-1OIS-1B-2B-2B-2N-1XV | 8% |
|  |  |  | 1B-1OIS-1B-2B-2OIS-2N-1XV | 7% |
| d2 | 46,241 | 0.8% | 1R-1OIS-1N-2S-2OIS-2NA-1XV | 87% |
|  |  |  | 1R-1OIS-1N-2R-2OIS-2NA-1XV | 4% |
|  |  |  | 1R-1OIS-1N-2R-2R-2NA-1XV | 2% |
| d3 | 81,464 | 1.4% | 1C-1OIS-1N-2S-2OIS-2NA-1XV | 91% |
|  |  |  | 1C-1OIS-1N-2S-2OIS-2N-1XV | 3% |
|  |  |  | 1C-1OIS-1R-2S-2OIS-2NA-1XV | 1% |
| d4 | 320,110 | 5.4% | 1ST-1OIS-1N-2B-2OIS-2N-1XV | 41% |
|  |  |  | 1ST-1OIS-1N-2R-2OIS-2N-1XV | 5% |
|  |  |  | 1ST-1OIS-1N-2B-2B-2N-1XV | 5% |
| d5 | 119,820 | 2.0% | 1A-1OIS-1N-2S-2OIS-2NA-1XV | 88% |
|  |  |  | 1A-1OIS-1N-2S-2OIS-2N-1XV | 7% |
|  |  |  | 1A-1OIS-1N-2A-2OIS-2NA-1XV | 1% |
| d6 | 39,166 | 0.7% | 2S-2OIS-2NA-1ST-1OIS-1N-2XV | 28% |
|  |  |  | 2B-2OIS-2N-1ST-1OIS-1N-2XV | 7% |
|  |  |  | 2ST-2OIS-2N-1ST-1OIS-1N-2XV | 5% |
| d7 | 205,907 | 3.5% | 1ST-1OIS-1B-2S-2OIS-2NA-1XV | 68% |
|  |  |  | 1B-1OIS-1B-2S-2OIS-2NA-1XV | 11% |
|  |  |  | 1ST-1OIS-1B-2S-2OIS-2N-1XV | 3% |
| d8 | 108,359 | 1.8% | 1ST-1OIS-1N-2ST-2OIS-2N-1XV | 72% |
|  |  |  | 1ST-1OIN-1N-2ST-2OIS-2N-1XV | 4% |
|  |  |  | 1ST-1OIS-1B-2ST-2OIS-2N-1XV | 4% |
| d9 | 72,006 | 1.2% | 1ST-1OIS-1N-2S-2OIS-2N-1XV | 86% |
|  |  |  | 1R-1OIS-1N-2S-2OIS-2N-1XV | 3% |
|  |  |  | 1E-1OIS-1N-2S-2OIS-2N-1XV | 2% |
| d10 | 87,269 | 1.5% | 1B-1OIS-1N-2S-2OIS-2NA-1XV | 87% |
|  |  |  | 1B-1OIS-1N-2S-2OIS-2N-1XV | 5% |
|  |  |  | 1B-1OIS-1N-2B-2OIS-2NA-1XV | 3% |
| d11 | 172,901 | 2.9% | 1ST-1OIS-1NA-2S-2OIS-2NA-1XV | 52% |
|  |  |  | 1A-1OIS-1NA-2S-2OIS-2NA-1XV | 18% |
|  |  |  | 1C-1OIS-1NA-2S-2OIS-2NA-1XV | 5% |
| d12 | 1,005,268 | 17.1% | 1ST-1OIS-1N-2S-2OIS-2NA-1XV | 89% |
|  |  |  | 1E-1OIS-1N-2S-2OIS-2NA-1XV | 1% |
|  |  |  | 1ST-1OIS-1N-2B-2OIS-2NA-1XV | 1% |

| Type | Weighted Count | % in Total | Representative Sequences | % in Type |
|---|---|---|---|---|
| e3 | 1,143 | 0.0% | 1ST-1OES-1BR-2ST-2OIS-2NA-1XV-1ROR-1XF-1NCH | 17% |
| | | | 1ST-1OIS-1N-2S-2OEN-2NA-1XV | 15% |
| | | | 1E-1OEN-1N-2ST-2OES-2N-1XV | 14% |
| f1 | 14,220 | 0.2% | 2ST-2ELL-2N-1ST-1OES-1N-2XV | 29% |
| | | | 2E-2ELL-2N-1ST-1OES-1N-2XV | 11% |
| | | | 2C-2ELL-2N-1C-1OES-1N-2XV | 6% |
| f2 | 23,919 | 0.4% | 1E-1OIS-1N-2S-2OIS-2NA-1XV | 16% |
| | | | 1ST-1OIS-1N-2S-2OIS-2NA-1XV | 11% |
| | | | 1PA-1OIS-1N-2S-2OIS-2NA-1XV | 6% |
| f3 | 20,576 | 0.3% | 2ST-2OES-2N-1E-1ELL-1N-2XV | 25% |
| | | | 2ST-2OES-2N-1E-1ERL-1N-2XV | 20% |
| | | | 2ST-2OES-2N-1E-1ERL-1NA-2XV | 5% |
| f4 | 32,310 | 0.5% | 1E-1ELL-1N-2S-2OES-2NA-1XV | 22% |
| | | | 1E-1ERL-1N-2S-2OES-2NA-1XV | 18% |
| | | | 1ST-1ERL-1N-2S-2OES-2NA-1XV | 9% |
| f5 | 8,931 | 0.2% | 1E-1ERL-1NA-2ST-2OES-2NA-1XV | 39% |
| | | | 1E-1ELL-1NA-2ST-2OES-2NA-1XV | 37% |
| | | | 1E-1OIS-1NA-2ST-2OES-2NA-1XV | 5% |
| f6 | 10,150 | 0.2% | 2ST-2ELL-2N-1S-1OES-1NA-2XV | 20% |
| | | | 2E-2OIS-2N-1S-1OIS-1NA-2XV | 8% |
| | | | 2ST-2OIS-2N-1S-1OIS-1NA-2XV | 5% |
| f7 | 18,779 | 0.3% | 1ST-1ERL-1N-2ST-2OES-2N-1XV | 25% |
| | | | 1ST-1N-1N-2ST-2N-2N-1XV | 16% |
| | | | 1ST-1OES-1N-2ST-2ELL-2N-1XV | 8% |
| f8 | 72,270 | 1.2% | 1E-1ELL-1N-2ST-2OES-2N-1XV | 67% |
| | | | 1E-1ELL-1NA-2ST-2OES-2N-1XV | 8% |
| | | | 1E-1ELL-1N-2C-2OES-2N-1XV | 4% |
| f9 | 99,879 | 1.7% | 1E-1ERL-1N-2ST-2OES-2N-1XV | 55% |
| | | | 1E-1ERL-1NA-2ST-2OES-2N-1XV | 8% |
| | | | 1PA-1ERL-1N-2ST-2OES-2N-1XV | 4% |

| Type | Weighted Count | % in Total | Representative Sequences | % in Type |
|---|---|---|---|---|
| g1 | 11,260 | 0.2% | 1ST-1ELL-1N-2ST-2OEO-2N-1XV | 18% |
| | | | 1ST-1ELL-1N-2ST-2OEO-2N-1CM-1XV | 12% |
| | | | 1C-1ELL-1N-2C-2OEO-2N-1CM-1XV | 7% |
| g2 | 2,918 | 0.0% | 1ST-1OET-1L-2S-2OEO-2NA-1XV | 10% |
| | | | 1ST-1N-1N-2ST-2N-2N-1XV | 10% |
| | | | 1C-1LCF-1N-2C-2OEO-2N-1XV | 7% |
| h1 | 586 | 0.0% | 1ST-1OEO-1L-2ST-2OEO-2N-1CM-1XV | 42% |
| | | | 1ST-1ELL-1N-2ST-2OEO-2N-1XV | 12% |
| | | | 1ST-1ELL-1L-2ST-2OEO-2N-1XV | 11% |
| i1 | 14,663 | 0.2% | 1ST-1ELL-1N-2S-2OEO-2NA-1XV | 10% |
| | | | 1C-1ELL-1N-2C-2OEO-2N-1XV | 7% |
| | | | 1C-1ELL-1N-2C-2OEO-2N-1CM-1XV | 5% |
| i2 | 3,405 | 0.1% | 2ST-2OEO-2N-1ST-1LCF-1N-2XV-1ROL-1XF | 16% |
| | | | 1ST-1OIO-1L-2ST-2OIO-2N-1XV | 14% |
| | | | 1ST-1LCO-1N-2ST-2OEO-2N-1XV | 8% |
| j1 | 89,654 | 1.5% | 1R-1R-1N-2ST-2OES-2N-1XV | 15% |
| | | | 1L-1ERL-1N-2L-2OES-2N-1XV | 8% |
| | | | 1R-1R-1N-2L-2OEO-2N-1XV | 7% |
| j2 | 544,349 | 9.2% | 1L-1L-1N-2ST-2OEO-2N-1XV | 52% |
| | | | 1L-1L-1N-2ST-2OES-2N-1XV | 5% |
| | | | 1L-1L-1NA-2ST-2OEO-2N-1XV | 4% |
| j3 | 461,634 | 7.8% | 2ST-2OEO-2N-1L-1L-1N-2XV | 51% |
| | | | 2ST-2ST-2N-1L-1OEO-1N-2XV | 6% |
| | | | 2ST-2ST-2N-1L-1L-1N-2XV | 5% |

| Type | Weighted Count | % in Total | Representative Sequences | % in Type |
|---|---|---|---|---|
| k1 | 50,571 | 0.9% | 1L-1ELL-1N-2S-2OEO-2NA-1XV | 18% |
|  |  |  | 1R-1ELL-1N-2S-2OEO-2NA-1XV | 17% |
|  |  |  | 1R-1R-1N-2S-2OEO-2NA-1XV | 14% |
| k2 | 82,579 | 1.4% | 2ST-2ST-2N-1L-1OET-1N-2XV | 61% |
|  |  |  | 2ST-2ST-2N-1R-1OET-1N-2XV | 5% |
|  |  |  | 2ST-2ST-2B-1L-1OET-1N-2XV | 4% |
| k3 | 228,645 | 3.9% | 2ST-2OEO-2N-1L-1L-1N-2XV | 43% |
|  |  |  | 2ST-2OES-2N-1L-1L-1N-2XV | 10% |
|  |  |  | 2ST-2ST-2N-1L-1L-1N-2XV | 7% |
| k4 | 183,944 | 3.1% | 1L-1L-1N-2ST-2OEO-2N-1XV | 57% |
|  |  |  | 1L-1L-1NA-2ST-2OEO-2N-1XV | 7% |
|  |  |  | 1L-1L-1N-2C-2OEO-2N-1XV | 5% |
| k5 | 21,835 | 0.4% | 1L-1L-1N-2S-2OEO-2NA-1XV | 49% |
|  |  |  | 1L-1L-1NA-2S-2OEO-2NA-1XV | 10% |
|  |  |  | 1L-1L-1N-2S-2OES-2NA-1XV | 8% |
| k6 | 175,550 | 3.0% | 1R-1R-1N-2ST-2OES-2N-1XV | 50% |
|  |  |  | 1R-1R-1NA-2ST-2OES-2N-1XV | 5% |
|  |  |  | 1R-1R-1N-2C-2OES-2N-1XV | 3% |
| k7 | 50,321 | 0.9% | 2ST-2OES-2N-1R-1R-1N-2XV | 50% |
|  |  |  | 2ST-2OES-2N-1R-1R-1NA-2XV | 6% |
|  |  |  | 2C-2OES-2N-1R-1R-1N-2XV | 5% |
| k8 | 26,569 | 0.5% | 1L-1OET-1N-2ST-2ST-2N-1XV | 53% |
|  |  |  | 1R-1OET-1N-2ST-2ST-2N-1XV | 6% |
|  |  |  | 1L-1OET-1N-2ST-2ST-2L-1XV | 4% |
| k9 | 190,854 | 3.2% | 1L-1L-1N-2ST-2OES-2N-1XV | 44% |
|  |  |  | 1L-1L-1NA-2ST-2OES-2N-1XV | 6% |
|  |  |  | 1L-1L-1N-2ST-2ST-2N-1XV | 5% |
| m1 | 91,068 | 1.5% | 1BU-1BU-1N-2S-2OIR-2NA-1XV | 40% |
|  |  |  | 1BU-1OIS-1N-2S-2OIR-2NA-1XV | 6% |
|  |  |  | 1BU-1BU-1NA-2S-2OIR-2NA-1XV | 6% |
| u | 116,156 | 2.0% | 1U-1U-1N-2ST-2OEO-2N-1XV | 6% |
|  |  |  | 1U-1U-1N-2ST-2OES-2N-1XV | 6% |
|  |  |  | 1ST-1N-1N-2ST-2N-2N-1XV | 4% |

| Type | Weighted Count | % in Total | Representative Sequences | % in Type |
|---|---|---|---|---|
| l1 | 12,508 | 0.2% | 2ST-2OET-2N-1ST-1ST-1N-2XV | 52% |
|  |  |  | 2ST-2OET-2N-1A-1ST-1N-2XV | 7% |
|  |  |  | 2ST-2OET-2B-1ST-1ST-1NA-2XV | 5% |
| l2 | 31,062 | 0.5% | 1A-1ST-1N-2ST-2OET-2N-1XV | 71% |
|  |  |  | 1A-1ST-1N-2A-2OET-2N-1XV | 5% |
|  |  |  | 1A-1ST-1N-2A-2ST-2N-1XV | 3% |
| l3 | 21,903 | 0.4% | 1ST-1OET-1N-2A-2ST-2N-1XV | 63% |
|  |  |  | 1ST-1OET-1B-2A-2ST-2N-1XV | 6% |
|  |  |  | 1ST-1OET-1N-2A-2OET-2N-1XV | 4% |
| l4 | 16,065 | 0.3% | 1ST-1OET-1N-2ST-2OET-2N-1XV | 85% |
|  |  |  | 1ST-1OET-1N-2ST-2OET-2N-1XV-2ROR-2XF | 2% |
|  |  |  | 1ST-1OET-1N-2S-2OET-2N-1XV | 1% |
| l5 | 10,319 | 0.2% | 1ST-1OET-1B-2ST-2ST-2N-1XV | 73% |
|  |  |  | 1ST-1OET-1B-2ST-2ST-2N-1XV-2RLO | 5% |
|  |  |  | 1ST-1OET-1B-2ST-2ST-2N-1XV-2ROL-2XF | 4% |
| l6 | 88,353 | 1.5% | 2ST-2ST-2N-1ST-1OET-1N-2XV | 54% |
|  |  |  | 2A-2ST-2N-1ST-1OET-1N-2XV | 7% |
|  |  |  | 2ST-2ST-2N-1ST-1ST-1N-2XV | 6% |
| l7 | 12,516 | 0.2% | 1ST-1ST-1N-2ST-2OET-2N-1XV-2ROR-2XF | 55% |
|  |  |  | 1ST-1ST-1N-2ST-2OET-2N-1XV-2ROR-2XF-2XF | 19% |
|  |  |  | 1ST-1ST-1N-2ST-2OET-2R-1XV-2ROR-2XF | 2% |
| l8 | 12,597 | 0.2% | 1ST-1ST-1NA-2ST-2OET-2NA-1XV | 47% |
|  |  |  | 1ST-1OET-1NA-2ST-2OET-2NA-1XV | 6% |
|  |  |  | 1ST-1ST-1NA-2ST-2ST-2NA-1XV | 5% |
| l9 | 43,371 | 0.7% | 1ST-1ST-1NA-2ST-2OET-2N-1XV | 56% |
|  |  |  | 1A-1ST-1NA-2ST-2OET-2N-1XV | 16% |
|  |  |  | 1ST-1ST-1NA-2ST-2OET-2B-1XV | 3% |
| l10 | 35,840 | 0.6% | 1ST-1OET-1N-2ST-2ST-2NA-1XV | 46% |
|  |  |  | 1ST-1OET-1N-2A-2ST-2NA-1XV | 10% |
|  |  |  | 1ST-1OET-1NA-2ST-2ST-2NA-1XV | 9% |
| l11 | 40,560 | 0.7% | 1ST-1ST-1N-2ST-2ST-2N-1XV | 85% |
|  |  |  | 1ST-1ST-1N-2ST-2ST-2N-1XV-1ROL-1XF | 2% |
|  |  |  | 1ST-1ST-1B-2ST-2ST-2N-1XV | 2% |
| l12 | 210,273 | 3.6% | 1ST-1OET-1N-2ST-2ST-2N-1XV | 75% |
|  |  |  | 1ST-1OET-1N-2ST-2ST-2N-1XV-2ROR-2XF | 2% |
|  |  |  | 1ST-1OET-1N-2ST-2ST-2N-1XV-2RLO | 2% |
| l13 | 385,833 | 6.6% | 1ST-1ST-1N-2ST-2OET-2N-1XV | 78% |
|  |  |  | 1ST-1ST-1N-2A-2OET-2N-1XV | 1% |
|  |  |  | 1ST-1ST-1N-2ST-2OET-2NA-1XV | 1% |
| l14 | 17,047 | 0.3% | 1ST-1ST-1B-2ST-2OET-2N-1XV | 61% |
|  |  |  | 1ST-1ST-1B-2ST-2OET-2NA-1XV | 5% |
|  |  |  | 1ST-1ST-1B-2ST-2OET-2B-1XV | 3% |